# HOT POINTS OF THE WAVE UNIVERSE CONCEPT: NEW WORLD OF MEGAQUANTIZATION

Chechelnitsky A.M.
Joint Institute for Nuclear Research, Laboratory of Theoretical Physics, 141980 Dubna, Russia

**ABSTRACT**

We present the brief review of some central problems, anomalies, paradoxes of Astrophysics, Cosmology and new approaches to its decisions, proposed by the Wave Universe Concept (WU Concept) (see monography - Chechelnitsky A.M. Extremum, Stability and Resonance in Astrodynamics ..., etc. and consequent publications).

All these, anyhow, - *Hot Points* of modern science of Universe, about which Standard Model have not authentic answers. The represented set of brief (justified) Novels reflects very perspective directions of search.

Essential from thems – Headline (and Topics):

# New Phenomenon – Megaquantization: Observed Megaquantum Effects in Astronimical Systems.

# Internal Structure of Celestial Bodies and Sun. Where is disposes the Convective Zone?

# Towards to Stars: Where the Heliopause will be found?

# Mystery of the Fine Structure Constant (FSC). Answer of the WU Concept: Theoretical Representation of the Fine Structure Constant. FSC – As Micro and Mega Parameter of the Universe. FSC, Orbits, Heliopause.

# What Quasars with Record Redshifts will be discovered in Future?

# Observed Universe: Monotonic Homogeneity and Anisotropy or Principal Hierarchy? Is the (Large – Scale) Limit of Universe exist?

All answers, proposed by the Wave Universe Concept, in contrast with speculative schemes of Standard Model, are effectively verified by experience, including, - and by justified in observations prognoses.

It will be waited, that namely here it is possible the real break in the understanding of these principal problems and enigmas, which Nature, infinite Universe offers (to us).

## MODERN COSMOLOGY: HOT POINTS – CHALLENGE WITHOUT ANSWER.

The manifold, unexpected, astonishing information about the structure of the Universe, represented by observations from Earth and Space, now is the real challenge to human Mind, to ability of the modern theory (in particular, of the Standard Model of Cosmology) adequately to understand the surrounding us World.

That challenge of observations and experiments, in many cases, remains without satisfactory Answer.

For the noneffective theory, which can not make the justified predictions (evidently, namely these are very essential attributes of the true science), every new significant phenomenon of observations and experiments is a suddenness, surprise, reason for astonishment. But obstinantly noninterpreting phenomena usually manifests a natural tendency to convert in anomalies, paradoxes, becomes Hot Points of science.

## CHALLENGE OF PROBLEMS.
## [But to Many Even Existing Answers there are not Yet Questions in the Modern Representations (in Standard Model)].

# Nothing or Something? Emptiness or Medium in Astrodynamics and Celestial Mechanics?

# In Search of Initial Principles.

What are Fundamental Laws, determining all variety of observed Micro and Mega Systems?

# What is commonness between Atom, Solar system, Galaxy?
Fundamental Isomorphism. Is it possible Unity and Universality of the dynamical description of the observed Hierarchy of Nature objects (systems)?
# Why Quantization? Hidden (Latent) sense of Quantization.
Quants of Microworld. Quantization “in the Small”. Dynamical Genesis. Planck Constant $\hbar$.
Quants of Megaworld. Megaquantization. Quantization “in the Large”. Megaquantum Effects. Dynamical Genesis. Constant $d^{-}$ (of Sectorial Velocity). Quantizations – is Pattern of only Microworld or of Megaworld and All Universe?
# Rhythms of Universe. Universal Phenomenon of periodicity. Why? Whence? From where?
Dynamical Genesis.
# Evolution. Whither is fly the “Arrow of Time”?
Direction of Evolution.
# Problems, Anomalies, Paradoxes of Modern Astrophysics, Cosmology (of Standard Model).

∗ Variable “Constant” of Hubble.
∗ Paradox of Missing Mass: Dark Matter – Phenomenon of Reality or only of the Theory.
∗ True of Microwave Background Radiation.
∗ Paradox of observed superlight velocities.
∗ Problem of the Redshift Quantization of far Qusars.

# What does the Universe look like on the very largest scales? How much Matter does it contain? What are the biggest Things that exist?
Etc, Etc…
# History of Universe and Man: Problems of Genesis.
*Genesis (Origin)* – by Myths, by Bible (by Moses), by Vedas;
*Genesis – by Modern Science*: Big Bang, Birth – Ex Nihilo.
# Earth and Universe – Problems and Paradoxies of Age.
Etc, Etc…
The List of these Problems, Anomalies, Paradoxes – is *nonterminating* (*Endless*).

**NEW PHENOMENON – MEGAQUANTIZATION:**
**Observed Megaquantum Effects in Astronimical Systems.**
**Wave Universe. Wave Astrodynamics.**

The principle possibility of dynamical synthesis of the classical celestial mechanics and astrophysics continual aspects, follows from the *Wave Universe concept* and wave concept of the astrodynamics - the *Megaquantum wave astrodynamics* (Chechelnitsky, 1980 -1997) (see Fig.1).

**Megawaves.** One of the basic ideas of the Wave Universe concept is the assertion of the existence of some *waves* in any megasystem (astronomical system) of Universe, in particular, in the Solar system. These waves actual realise short-range interactions in the scale, compared with scale of the system.

**The Suggestion.** Let us study the Solar system, - as the *wave dynamic system* (WDS)of megaworld. There are *megawaves* (large astronomical scale waves with large lengths and periods), which induce, propagate, and absorbe due to the *cosmic medium*. These megawaves are responsible for the dynamic structure (and geometry) of the Solar system (the *co - dimension principle* (Chechelnitsky, 1980)).

**Micro-Mega Analogy.**

When quantum mechanics was developing, the main role in the comprehension of the microobject dynamic structure belonged to the analogy of the structure between the atom and planetary system.

Apparently, the time has come to return debts. And at present this dynamic analogy works vise versa: now astronomical (in particular, planetary) systems are studied in analogy - with atom systems (Fig. 1).

Besides the investigation of *Fundamental wave equations* (Chechelnitsky, 1980, 1986), some quantitive representations on megawaves properties (if these waves are considered as some *analog* *of the De Broglie waves for megaworld astronomical systems)* can be received, in particular, from the following relations of the Megaquantum wave astrodynamics:

$$v = \bar{d}\cdot k \quad \varepsilon = \bar{d}\cdot\Omega \quad \Delta x\cdot\Delta v \geq (1/2)\cdot\bar{d}$$

Here $k$ - wave number, $v$ - velocities, $\Omega$ - circular frequency,

$\varepsilon$ - normalized (divided by m -mass) energy of megawaves ($\varepsilon = E/M=(1/2)\cdot v^2$),

$\bar{d} = d/2\pi$ - Fundamental constant of normalized action (sectorial velocity, circulation) with the dimension [$cm^2\cdot s^{-1}$].

The quantization constant $\bar{d} = d/2\pi$ in the microworld for the atom is determined by Planck's constant $\hbar=h/2\pi$ and electron mass $m_e$

$$\bar{d} \sim \hbar/m_e = 1.158\ cm^2\cdot s^{-1}$$

The mentioned relations above represent the megaquantum analogs of the *Bohr, Planck-Einstein, Heisenberg relations* well-known in the microworld quantum mechanics, accordingly. It should be emphasized that, at least, in the frame of the Wave astrodynamics, the formal analog of Heisenberg incertainty relation ($\Delta x\cdot\Delta v \geq (1/2)\bar{d}$ - *difractional* incertainty relation) does not have such a wide prohibitive sense, as in the Copenhagen interpretation of microworld quantum mechanics (Born, 1963, Jammer, 1985).

**Megaquantum World of the Solar System Distance Quantization.**

**Radial Quantization.** The high-precision information on the geometry (in particular, on semi-major axes) of planetary orbits, gives the possibility to receive some single-valued, intriguing facts. They were found in the 70-ies using the quantum of (linear size) distance $a_* = a_*^{[1]} = 8R_\odot$ ($R_\odot$ - radius of Sun) and caused the astonishment surprise showing a great set of *integer* numbers, which characterized the dynamical structure and geometry of the Solar system (Fig. 2).

It concerns differences of $\Delta a\hat{}$ - normalized (divided by $a_*$) planetary distances $a\hat{} = a/a_*$ for orbits $a_i$, $a_j$

$$\Delta a\hat{}_{i,j}=a\hat{}_i - a\hat{}_j, \quad (i, j = 1,2,3....),$$

as however, in many cases, and on the normalized distances $a\hat{}_i = a_i / a_*$ themselves.

The integerality phenomenon

$$\Delta a\hat{} = \text{Integer [Semi-Integer]},$$

as and the $a\hat{}$ = Integer [Semi-integer],

has a deep physical basis. Such a nonaccidental abundance of integer (semi-integer) numbers (in the absence of anything ad hoc fitting parameters) shows the conceal, early unknown, but really existing phenomenon of the Solar system *wegawave* structure (Chechelnitsky, 1984; 1992 b, c, Table 1) (see Fig. 3).

**Azimuthal Quantization.** The effects of *megaquantization (quantization "in the Large") - the megaquantum effects*, are not less interesting. They were discovered for $p_i = 2\pi a_i$ perimeters of planetary orbits, normalized by $a_*$ quantum, i.e. for

$$p\hat{}_i = p_i / a_* = 2\pi a_i / a_* = 2\pi a\hat{}_i ,$$

$$\Delta p\hat{}_{i,j} = p\hat{}_i - p\hat{}_j$$

Azimuthal Quantization $p\hat{}$, $\Delta p\hat{}$ = Integer [Semi-Integer]

is a very characteristic pattern of the physically distinguished orbits and it occurs rather often (Chechelnitsky, 1992 b,c, Table 1) (see Fig.3A).

**Sectorial Velocity Quantization.** In this case we speak about the observed effect of the discreteness - the quantization of the dynamical value

$$L_N = (K\cdot a)^{1/2} = L_{N=1}\cdot N$$

of the *sectorial velocity*, normalized by

$$L_{N=1} = L_*/(2\pi)^{1/2}, \text{ where } L_* = (K\cdot a_*)^{1/2}.$$

Taking into account the interpretation of N quantum number, as $N = L_N/L_{N=1}$, we can talk about integer (semiinteger) ability of the N quantum number.

For planetary orbits of Mercury (ME), Venus (V), Earch (E), Mars (MA), we have, in particular, $N=(2\pi a/a_*)^{1/2}$

N = 7.911; 11.050; 12.99; 15.969, close to *integer*

N = 8; 11; 13; 16, accordingly (Fig. 2) (see also Fig. 3).

At the definite conceptual supposition this effect may be connected with the well-known quantization effect of kinetical momentum ($K^{(m)} = mva$, where m - mass) or action.

In this aspect it has been known since the times Planck, Einstein, Bohr, De Broglie. Although, we should stress, that in the quantum mechanics (of microworld) the quantization namely of the kinetical momentum $K^{(m)} = mva$ was always discussed, but not of the *sectorial velocity (or circulation)* $L = K^{(m)}/m = va$.

Since Kepler's times (his second law) the notion and dynamical value of the *sectorial velocity* has taken the importance place in the astrodynamics, in space sciences.

**INTERNAL STRUCTURE OF CELESTIAL BODIES AND SUN.**
**Where is disposes the Convective Zone?**

Results from Helioseismology, periodicities spectrum of the Sun and stars - can be adequately understand and effectively interpreted only in the context of unified theoretical representations. That possibility is permitted the *Wave Universe concept* (see monograph - Chechelnitsky A.M. Extremum, Stability, Resonance in Astrodynamics ..., etc. and consequent publications), which comprehand the Solar and any astronomical systems as the principle *Wave dynamic systems* (WDS). Investigations with using of the theoretically calculated *Fundamental Wave spectrum of (periods)* those objects, its *Megaspectroscopy* (Fig. 4) (including the Helioseismology - for the Sun, and Asteroseismology - for the stars) lead to reconstruction of dynamics, physics, geometry of the Sun and stars (its internal and external structure) (Fig. 5).

The Hierarchy set of $TR_*^{[s]}$ (s = …, -2, -1, 0, 1, 2, …) Transspheres of the Solar system with semi-major axes

$$a_*^{[s]} = \chi^{2(s-1)} a_*^{[1]} = \chi^{2(s-1)} 8R_\odot, \quad s = \ldots, -2, -1, 0, 1, 2, \ldots,$$

it is clear, also contain the value of critical surface - $TR_*^{[0]}$ Transsphere

$$a_*^{[0]} = \chi^{-2} 8R_\odot = 0.595R_\odot,$$

which lies inside the Sun and coincide with theoretical (geometrical) *beginning of the Convective Zone* (with the boundary of External Nucleus) of the Sun:

| | |
|---|---|
| Internal (Smallest) Nucleus of Sun | …, $a_*^{[-1]} = \chi^{-4} 8R_\odot = 0.0442R_\odot$, |
| External Nucleus of Sun (Beginning of Convective Zone) | $a_*^{[0]} = \chi^{-2} 8R_\odot = 0.595R_\odot$, |
| $TR_*^{[1]}$ Transsphere (for I Earth Group of Planets) | $a_*^{[1]} = 8R_\odot = 0.0372193$ AU, |
| $TR_*^{[2]}$ Transsphere (for II Jupiter Group of Planets) | $a_*^{[2]} = \chi^{2} 8R_\odot = 0.500$ AU, |
| $TR_*^{[3]}$ Transsphere (for Trans-Pluto Group of Objects) | $a_*^{[3]} = \chi^{4} 8R_\odot = 6.727$ AU, |
| $TR_*^{[4]}$ Transsphere – Heliopause | $a_*^{[4]} = \chi^{6} 8R_\odot = 90.447$ AU, |
| $TR_*^{[5]}$ Transsphere | $a_*^{[5]} = \chi^{8} 8R_\odot = 1216.016$ AU,… , |

$\chi$ - *Fundamental parameter of hierarchy - Chechelnitsky Number* $\chi = 3.66(6)$.

**TOWARDS TO STARS: WHERE THE HELIOPAUSE WILL BE FOUND?**
**Where the Solar System Ends?**

Of course, such a problem is specially attracting and interesting, when spacecrafts Pioneer 10, 11 and Voyager 1,2 still did't achieve the intriguing, significant barrier of the Solar system - *Heliopause [Abstracts COSPAR, 1994; Belcher et al., 1993; Masek, 1996]*.

Some astrophysicists are strive to consider the Heliopause as objectively detectable external boundary of the Solar system. But astronomers - observers will hardly agree with this conclusion. Resulting from these and other, more fundamental ideas, it will be reasonable to suppose, that the region beyond Heliopause also has very nontrivial properties. Just from the fact that for a long time it has been considered as inexhaustable reservoure of comet bodies.

**Near the Stars. Trans - Pluto Space. Heliopause Region**

The prognosis, based upon the conceptions, besides the existence regions - $G^{[1]}$, that is occupied by the space of I (Earth) group planets, and $G^{[2]}$, that is occupied by the space of II (Jupiter) group, predicts the existens of at least, Trans - Pluto $G^{[3]}$ Shell and, probably, the following it $G^{[4]}$ Shell [Chechelnitsky, 1992 a, b, c] (Fig. 6).

***$G^{[3]}$ Shell.*** The *dominant orbits* system in this Trans-Pluto Shell with using of the (linear size) distance quantum

$$a_*^{[3]} = K_\odot/( C_*^{[3]})^2 = 6.7276 \text{ AU}$$

in the main approach are as follow ($a^{[s]}=a_*^{[s]}N^2/2\pi$, $K_\odot$ - gravitational parameter of Sun):

$$a^{[3]} = 68.5(70); 90.44; 129.5; 181; 257; 274; 407(403); 495(500); 542(530) \text{ AU}$$

***$G^{[4]}$ Shell.*** In this far region of the Solar system with the usage of the (linear size) distance quantum

$$a_*^{[4]} = K_\odot/(C_*^{[4]})^2 = 90.447 \text{ AU}$$

the geometry of *dominant orbits* (semi-major axes values) seem as following

$$a^{[4]} = 921; 1216; 1742; 2433; 3458; 3680; 5474; 6654; 7287 \text{ AU}$$

**Heliosphere Boundaries. Heliopause**

According to the ideas of modern astrophysics, the interaction of the Sun with interstellar medium, surrounding it, results in formation of the Solar system *Heliosphere* (that in some sence reminds the well investigated Earth *magnetosphere*). Depending on the chosed parameters, it is believed that the bow shock, after which the Heliosphere begins, can be located at the distance 30 - 50 AU or 75 - 200 AU.

There are reasons to expect, that the bow shock (and inside it - the *Heliopause*) may be discovered at the heliocentric distance [Chechelnitsky, 1992 a,b,c]

$$a_* = a_*^{[4]} = K_\odot/(C_*^{[4]})^2 = 90.447 \text{ AU},$$

connected with the Solar system $G^{[4]}$ Shell (Fig. 6).

**Trans - Pluto Celestial Bodies**

The potentially possible Trans - Pluto celestial bodies most probably can be discovered on these dominant orbits.

The experience of space researches of the Solar system and satellite systems of planets (not depending on the succeses in far celestial bodies continuating search) show, that dominant orbits are physically distinguished states in many other, fixed by observations, aspects (for instance, - in measurements of energetic proton count intensity, etc.) [Chechelnitsky, 1992 a].

**MYSTERY OF THE FINE STRUCTURE CONSTANT (FSC).**

**Microworld: Quantum Wave Mechanics and Fine Structure Constant**

From all modern theories of microworld - quantum electrodynamics (QED) describes the dynamic structure and the interaction of elementary particles (photons, electrons, muons) most exactly.

There is the fundamental parameter (coupling constant, interaction parameter), that lies in the basis of that advanced and consistent theory - the Fine Structure Constant (FSC), [Born, 1963].

The theoretical representation of this constant is unknown up till now. "The Mysterious Number 137" - so titled Max Born the famous paper of 1936 [Born, 1936].

The Fine Structure Constant (FSC) $\alpha = 2\pi e^2/hc$ or nondimensional number $\alpha^{-1} \approx 137$ (where e - electron charge, h - Planck constant, c - speed of the light) was introduced in the theoretical physics by Arnold Sommerfeld in 1915 [Sommerfeld, 1973]. That is the fundamental parameter of the all atomic spectroscopy. At present, only its experimental value is known ($\alpha^{-1} = 137.036$).

**Answer of the WU Concept:**
**Theoretical Representation of the Fine Structure Constant.**

In the framework of Wave Universe concept may be naturally obtained the following surprisely simple analytical and numerical (closed) representation for the Fine Structure Constnt that is proved to be correct by the logics of the consistent theory [Chechelnitsky, (1986) 1996]

$$\alpha^{-1} = 2^{39/4}/2\pi = 137.0448088$$

**FSC – As Micro and Mega Parameter of the Universe.**
**FSC, Orbits, Heliopause.**

Fine Structure Constant is fundamental constant not only of microworld (atoms), but *also – of megaworld* (astronomical systems) – one of the general nondimensional parameter of Universe.

**Megaworld: Megaquantum Wave Astrodynamics and Astrophysics;**
**Earth Orbit and Heliopause**

We shall cite only one fragment of the new knowledge [Chechelnitsky, 1996], that is spontaneously connected with discussed theme - with wave structure, geometry and dynamics of Solar system.

**The Assertion**

∗ There is regular connection between *planetary orbits* arrangement and special *critical surface* of Solar system - Heliopause location.

∗ This connection may be presented by using the Fine - Structure Constant, that is considered as *megaparameter* of astronomical systems.

∗ In particular, when using the Earth orbit, the following most simple relation between the Keplerian periods of Earth orbit $T_E = 1^a$ and of Heliopause $T_*$ is valid:

$$T_* \approx S_\alpha T_E \approx 861^a, \quad S_\alpha = 2\pi/\alpha \approx 2\pi \cdot 137 \approx 861$$

The appearing from the above relation between semi-major axis of Earth orbit $a_E$ = 1 AU and of Heliopause $a_*$ is like this:

$$a_* \approx S_\alpha^{2/3} \cdot a_E = 90.5 \text{ AU}$$

These relations reflect the presence of spontaneous and close connection between Wave astrodynamics (celestial mechanics) - geometry and dynamics of regular set of elite (dominante) Solar system planetary orbits - and geometry and dynamics of Heliopause (of Solar system magnetosphere, or of standing shock wave of Heliosphere), that is traditionally regarded as an object of astrophysics.

**WHAT QUASARS WITH RECORD REDSHIFTS WILL BE DISCOVERED IN FUTURE?**
**Megaquantization in the Universe.**

It is clear, *Megaquantization* (quantization “in the Large”), observed megaquantum effects are not monopolic privelege of only Solar system.

Let us point the brief resume of research (prognosis), connected with problem of redshift quantization of far objects of Universe – quasars (QSO) [Chechelnitsky, (1986) 1977]:

“*Abstract*: In the framework of the Wave Universe concept it is shown that the genesis of redshifts can be connected with the *intra-system (endogenous)* processes which take place in astronomical systems. The existence of extremal redshift objects (quasars – QSO) with most probable

$$z = 3.513\ (3.847);\ 4.677;\ 6.947\ (7.4);\ 10.524;\ 14.7;\ 27.79;\ \ldots$$

is predicted.”

Prognosis already had justified successively for extremal values of z redshifts

$z_{theory} = 3.513$, $z_{obs} = 3.53$ (quasar OQ172)
$z_{theory} = (3.847)$, $z_{obs} = 3.78$ (quasar PKS2000-330)
$z_{theory} = 4.677$, $z_{obs} = 4.71$ (Schmidt, Gunn, Schnaider, 1989)
$z_{obs} = 4.694$ (4.672) (quasar BR1202-0725, Wampler et al., 1996)

At the present time, apparently, also the object Q2203+29 G73 with record value z of redshift z=6.97 is discovered in special Astrophysical Observatory (SAO, Russia)

$$z_{theory} = 6.947, \quad z_{obs} = 6.97$$

(Q2203+29 G73, Dodonov et al., 2000).

The Quene – for objects with even more high redshifts z = 10.524; 14.7; …

Consequences of such successfully realizable prognosis, imperatives of observations not only are unexpected for the Standard cosmology, but also, probably, its can stimulated the radical reconsideration of many habitual representations, having become as freezen dogmas.

## OBSERVED UNIVERSE:
## MONOTONIC HOMOGENEITY AND ANISOTROPY OR PRINCIPAL HIERARCHY?
### Is the (Large – Scale) Limit of Universe exist?

Invariablly justified representations of the Wave Universe Concept - WU Concept indicate a principle incorrectness of expectations of Standard (Model) cosmology about homogeneity and isotropy of the Universe.

It also is connected with observational data about apparent hierarchy of giant astronomical systems (stellar systems, galaxies, clusters of galaxies, superclusters of galaxies, etc.), their megawave structure, Megaquantization (quantization "in the Large"), non-homogeneity of microwave background Space radiation, adequately interpreted (in frameworks of WU Concept) effects redshifts quantization of quasars, etc.

The principle absence of a Limit of Hierarchy of Matter Levels asserts: "The Staircase of a Matter" - is endless.

For orientation of the explorers, working with the observational data, in frameworks of WU Concept the concrete characteristics of following (behind superclusters of galaxies) potentially possible extremely large astronomical systems are calculated with using the Fundamental parameter of Hierarchy – Chechelnitsky Number $\chi$ = 3.66(6).

The astronomical systems, belonging to the nearest hierarchy Levels of Solar-Like systems, are characterized by external radiuses [$a^{(k)} = \chi^k a^{(0)}$, $a^{(0)}$ = 39.373 AU]

$$a^{(20)} = 36.83,\ a^{(21)} = 135,\ a^{(22)} = 495,\ a^{(23)} = 1815 \text{ Mpc}.$$

It may be expected that in the Universe also exist and should show itself in observations (the Solar-Like objects) – extremely large astronomical systems (ELAS), characterized by the external radiuses (of peripherals)

$$a^{(26)} = 89503,\ a^{(27)} = 328177,\ a^{(28)} = 1203318 \text{ Mpc}.$$

## FROM PARADOXES AND ANOMALIES – TO THE GREAT SYNTHESIS.
### Foreseeing Future.

Many factors, - including megaquantum effects, manifestation of the FSC as megaparameter in the science of giant astronomical systems – (Mega) Wave astrodynamics, in our opinion, are symptoms, that in front of sciences about Universe new, (for many, possibly) surprisely representatives are opening:

Sciences about Universe are found at the Way to the Great Synthesis [Chechelnitsky, 1996-1997], in particular, of Substance and Medium.

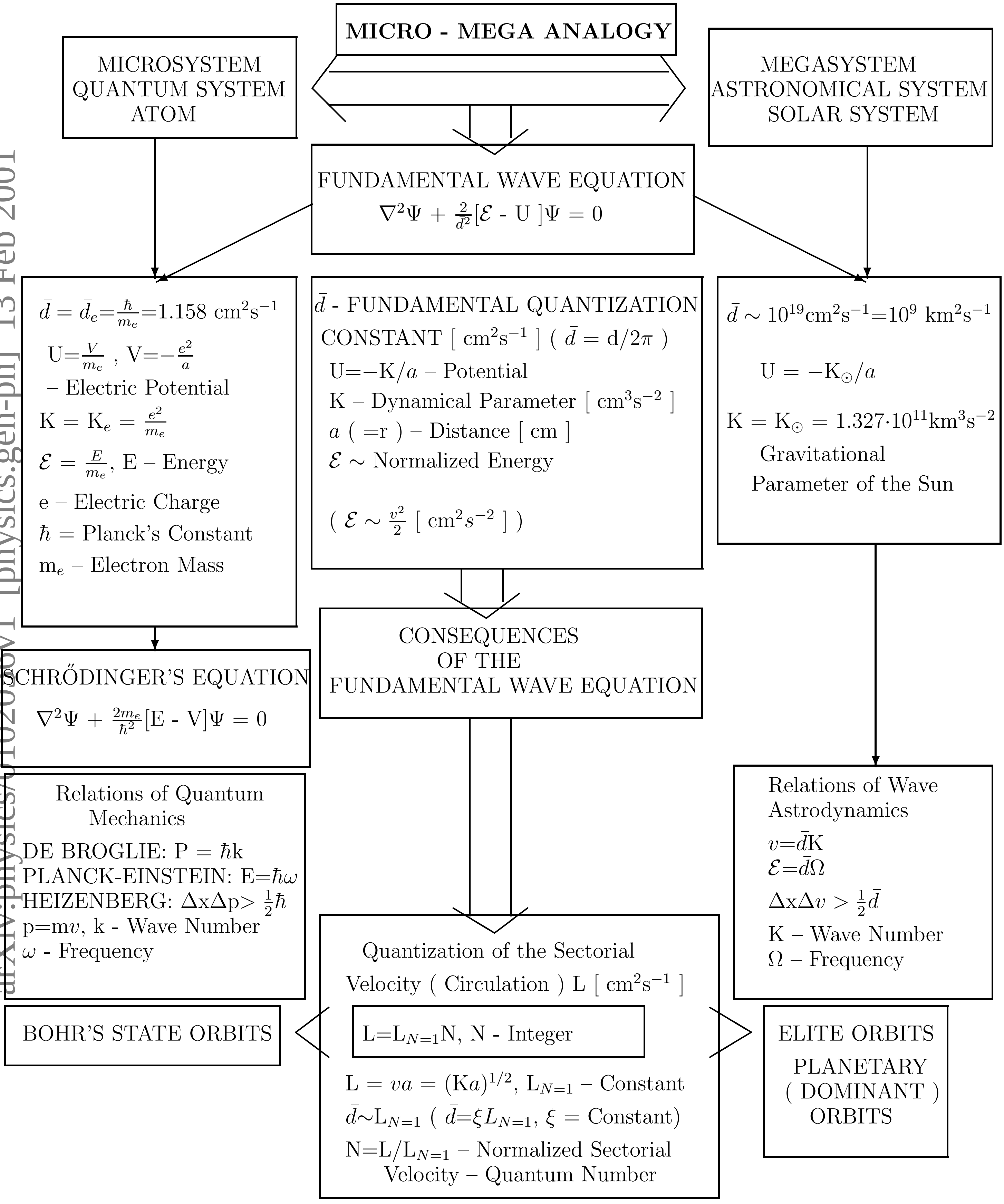
MICRO - MEGA ANALOGY
MICROSYSTEM QUANTUM SYSTEM ATOM
MEGASYSTEM ASTRONOMICAL SYSTEM SOLAR SYSTEM
FUNDAMENTAL WAVE EQUATION
$\nabla^2\Psi + \frac{2}{\bar{d}^2}[\mathcal{E} - U]\Psi = 0$
$\bar{d} = \bar{d}_e = \frac{\hbar}{m_e} = 1.158$ cm$^2$s$^{-1}$
$U = \frac{V}{m_e}$ , $V = -\frac{e^2}{a}$
– Electric Potential
$K = K_e = \frac{e^2}{m_e}$
$\mathcal{E} = \frac{E}{m_e}$, E – Energy
e – Electric Charge
$\hbar$ = Planck's Constant
$m_e$ – Electron Mass
$\bar{d}$ - FUNDAMENTAL QUANTIZATION CONSTANT [ cm$^2$s$^{-1}$ ] ( $\bar{d}$ = d/2$\pi$ )
U=−K/$a$ – Potential
K – Dynamical Parameter [ cm$^3$s$^{-2}$ ]
$a$ ( =r ) – Distance [ cm ]
$\mathcal{E}$ ∼ Normalized Energy
( $\mathcal{E} \sim \frac{v^2}{2}$ [ cm$^2 s^{-2}$ ] )
$\bar{d} \sim 10^{19}$cm$^2$s$^{-1}$=$10^9$ km$^2$s$^{-1}$
$U = -K_\odot/a$
$K = K_\odot = 1.327\cdot10^{11}$km$^3$s$^{-2}$
Gravitational Parameter of the Sun
SCHRŐDINGER'S EQUATION
$\nabla^2\Psi + \frac{2m_e}{\hbar^2}[E - V]\Psi = 0$
CONSEQUENCES OF THE FUNDAMENTAL WAVE EQUATION
Relations of Quantum Mechanics
DE BROGLIE: P = $\hbar$k
PLANCK-EINSTEIN: E=$\hbar\omega$
HEIZENBERG: $\Delta x \Delta p > \frac{1}{2}\hbar$
p=m$v$, k - Wave Number
$\omega$ - Frequency
Relations of Wave Astrodynamics
$v = \bar{d}K$
$\mathcal{E} = \bar{d}\Omega$
$\Delta x \Delta v > \frac{1}{2}\bar{d}$
K – Wave Number
$\Omega$ – Frequency
Quantization of the Sectorial Velocity ( Circulation ) L [ cm$^2$s$^{-1}$ ]
L=$L_{N=1}$N, N - Integer
L = $va$ = (K$a$)$^{1/2}$, $L_{N=1}$ – Constant
$\bar{d} \sim L_{N=1}$ ( $\bar{d} = \xi L_{N=1}$, $\xi$ = Constant)
N=L/$L_{N=1}$ – Normalized Sectorial Velocity – Quantum Number
BOHR'S STATE ORBITS
ELITE ORBITS
PLANETARY ( DOMINANT ) ORBITS

Figure 1.

# MEGAQUANTUM EFFECTS IN THE SOLAR SYSTEM

## INITIAL DATA

## QUANTIZATION "IN THE LARGE": QUANTIZATION OF DISTANCIES ( Linear Values )

* The Astronomical Ephemeris DE19 JPL; $a$-Semi-Major Axes of Planetary Orbits.
* The radius of the Sun $R_\odot = 695992$ km ( $\approx 696000$ km )
* $a = a_* = 8R_\odot = 5567928$ km $= 0.0372193$ AU

Semi-Major Axes of the Transsphere

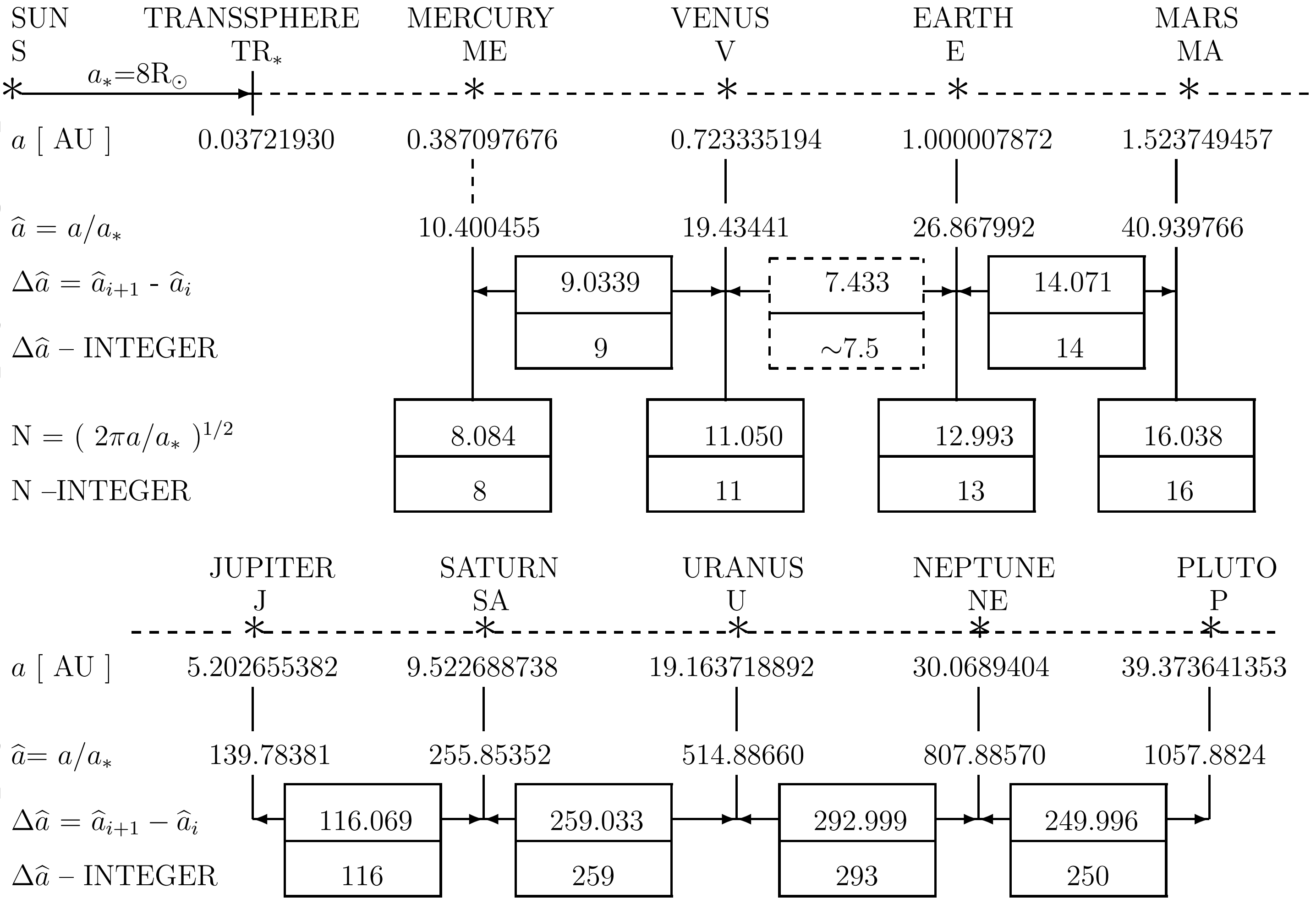

Figure 2.

Fig. 3

# MEGAQUANTUM EFFECTS IN THE SOLAR (PLANETARY) SYSTEM
# RADIAL, AZIMUTHAL, SECTORIAL QUANTIZATION

**(** - Close to Integer, * - Close to Semi-Integer)**

| Shells Planetary Groups | Planets and Orbits | | Semi-Major Axes of the Planetary Orbits (Astronomical Ephemeris DE19 JPL | | | $G^{[1]}$ Shell $C_*^{[1]} = 154.3864$ km·s$^{-1}$ $a_*^{[1]} = 0.0372$ AU = 8R | | | $G^{[2]}$ Shell $C_*^{[2]} = 42.1$ km·s$^{-1}$ $a_*^{[2]} = 0.5$ AU | | | $G^{[3]}$ Shell $C_*^{[3]} = 11.4$ km·s$^{-1}$ $a_*^{[3]} = 6.727$ AU | | |
|---|---|---|---|---|---|---|---|---|---|---|---|---|---|---|
| | Orbits (Planets) | In-dex | a [AU] | a [km] | a/R R = =695992 km | Radial Quantization $\hat{a}=a/a_*$ $\Delta\hat{a}=\hat{a}_i-\hat{a}_j$ | Azimuthal Quantization $\hat{P}=2\pi\hat{a}$ $\Delta\hat{P}$ | Sectorial Quantization $N=(2\pi\hat{a})^{1/2}$ $\Delta N$ | a $\Delta\hat{a}$ | P^ $\Delta\hat{P}$ | N $\Delta N$ | a $\Delta\hat{a}$ | P^ $\Delta\hat{P}$ | N $\Delta N$ |
| I | Mercury | ME | 0.3871 | 57908887. | 83. | 10.400 | 65.348 | ** 8.084 | 0.774 | 4.861 | 2.205 | 0.058 | 0.362 | 0.601 |
| | | | | | | ** 3.044 | 19.126 | | | | | | | |
| (Earth) | | TR | 0.5004 | 74857428. | 108. | * 13.444 | * 84.474 | 9.192 | ** 1.000 | 6.283 | * 2.507 | 0.074 | * 0.467 | 0.684 |
| | | | | | | ** 5.990 | | | * 0.446 | | | | | |
| | Venus | VE | 0.7233 | 108209375. | 155. | * 19.434 | 122.110 | ** 11.050 | * 1.446 | ** 9.083 | **3.014 | 0.108 | 0.676 | 0.822 |
| Planetary | | | | | | * 17.434 | | ** 1.943 | * 0.552 | | | | | |
| | Earth | EA | 1.0000 | 149598917. | 215. | 26.868 | 168.816 | ** 12.993 | ** 1.998 | ** 12.557 | * 3.544 | 0.149 | **0.934 | **0.966 |
| Group | | | | | | ** 14.072 | * 88.416 | ** 3.045 | ** 1.047 | | | | | |
| | Mars | MA | 1.5237 | 227949605. | 328. | ** 40.940 | 257.232 | ** 16.038 | ** 3.045 | 19.133 | 4.374 | 0.226 | * 1.423 | 1.193 |
| | | | | | | * 33.417 | ** 209.964 | * 5.577 | * 2.486 | | * 1.521 | | | |
| | Ceres | CE | 2.7675 | 414012105. | 595. | 74.357 | 467.196 | 21.615 | * 5.531 | 34.750 | 5.895 | * 0.411 | * 2.585 | 1.608 |
| | | | | | | | | | | | | | | |
| II | Jupiter | JP | 5.2027 | 778306106. | 1118. | 139.781 | 878.288 | 29.636 | 10.397 | 65.327 | ** 8.083 | 0.773 | 4.859 | 2.204 |
| | | | | | | ** 40.968 | | | ** 3.047 | 19.147 | | | | |
| (Jupiter) | | TR | 6.7275 | 1006412639. | 1446. | 180.752 | 1135.697 | 33.700 | * 13.444 | * 84.474 | 9.191 | **1.000 | 6.283 | * 2.507 |
| | | | | | | 75.102 | | | | ** 35.098 | | 0.415 | | * 0.475 |
| | Saturn | SA | 9.5227 | 1424573841. | 2047. | 255.854 | * 1607.575 | ** 40.095 | ** 19.030 | * 119.572 | ** 10.935 | * 1.415 | 8.891 | ** 2.982 |
| Planetary | | | | | | ** 259.033 | | | | **121.058 | | * 1.434 | ** 9.004 | |
| | Uranus | UR | 19.1637 | 2866851394. | 4119. | 514.887 | 3535.128 | 56.878 | 38.297 | 240.630 | * 15.512 | 2.849 | 17.898 | 4.231 |
| | | | | | | ** 292.999 | | | | **136.932 | | | | ** 1.068 |
| Group | Neptune | NE | 30.0689 | 4498249377. | 6463. | 807.886 | ** 5076.096 | 71.247 | ** 60.091 | * 377.362 | * 19.431 | * 4.470 | ** 28.083 | 5.299 |
| | | | | | | ** 249.997 | | | | | | | | |
| | Pluto | PL | 39.3736 | 5890212828. | 8463. | 1057.883 | 6646.872 | * 81.528 | 78.686 | 494.397 | 22.235 | 5.853 | 36.773 | ** 6.064 |

| Mercury Series | Venus Series | Earth Series | Mars Series | Ceres Series (Asteroids) | Jupiter Series | Saturn Series | Uranus Series | Neptune Series | Pluto Series |
|---|---|---|---|---|---|---|---|---|---|
| 58,$^{d}$6809 | | | | | | | | | |
| 44,505 | | | | | | | | | |
| 35,567 | | | | | | | | | |
| 31,713 | | | | | | | | | |
| 29,470 | | | | | | | | | |
| 23,43 | | | | | | | | | |
| 27,840 | | | | | | | | | |
| 27,633 | | | | | | | | | |
| 27,548 | | | | | | | | | |
| 27,$^{d}$277 | | | | | | | | | |
| | 184,$^{d}$229 | | | | | | | | |
| | 97,033 | | | | | | | | |
| | 69,007 | | | | | | | | |
| | 59,201 | | | | | | | | |
| | 55,160 | | | | | | | | |
| | 52,970 | | | | | | | | |
| | 52,227 | | | | | | | | |
| | 51,924 | | | | | | | | |
| | 50,$^{d}$970 | | | | | | | | |
| | | 205,$^{d}$013 | | | | | | | |
| | | 110,335 | | | | | | | |
| | | 87,234 | | | | | | | |
| | | 78,735 | | | | | | | |
| | | 74,346 | | | | | | | |
| | | 72,891 | | | | | | | |
| | | 72,303 | | | | | | | |
| | | 70,$^{d}$467 | | | | | | | |
| | | | 238,$^{d}$919 | | | | | | |
| | | | 151,845 | | | | | | |
| | | | 127,827 | | | | | | |
| | | | 116,648 | | | | | | |
| | | | 113,104 | | | | | | |
| | | | 111,695 | | | | | | |
| | | | 107,$^{d}$373 | | | | | | |
| | | | | 416,$^{d}$647 | | | | | |
| | | | | 274,911 | | | | | |
| | | | | 227,932 | | | | | |
| | | | | 214,784 | | | | | |
| | | | | 209,759 | | | | | |
| | | | | 195,$^{d}$016 | | | | | |
| | | | | | 808,$^{d}$128 | (2,$^{a}$212) | | | |
| | | | | | 503,233 | (1,$^{a}$377) | | | |
| | | | | | 443,317 | (1,$^{a}$213) | | | |
| | | | | | 422,431 | (1,$^{a}$156) | | | |
| | | | | | 366,$^{d}$613 | (1,$^{a}$00375) | | | |
| | | | | | | 3,$^{a}$651 | | | |
| | | | | | | 2,688 | | | |
| | | | | | | 2,423 | | | |
| | | | | | | 1,$^{a}$837 | | | |
| | | | | | | | 10,$^{a}$194 | | |
| | | | | | | | 7,203 | | |
| | | | | | | | 3,$^{a}$697 | | |
| | | | | | | | | 24,$^{a}$548 | |
| | | | | | | | | 5,$^{a}$801 | |
| | | | | | | | | | 7,$^{a}$596 |

Figure 4. Fundamental Wave Spectrum of the Solar System ( Fragment )
( $\mathcal{T}$ – Wave Periods [ in d-days, a-years ] )

**Fig. 5**

# INTERNAL STRUCTURE OF CELESTIAL BODIES (Fragment: Sun, Earth, Moon)

## Isomorphysm. Dominant Levels (States)

| Level Index | Dominant (Planetary) Values N | $a^\wedge = a/a_* = N^2/2\pi$ | Sun (R=695992 km) | | Earth ($R_{mid}$=6371km) | | | | | | Moon (R = 1738 km) | | | | | | | |
|---|---|---|---|---|---|---|---|---|---|---|---|---|---|---|---|---|---|---|
| | | | $G^{[-2]}$ Shell | $G^{[-1]}$ Shell | $G^{[1]}$ Shell | | | $G^{[2]}$ Shell | | | $G^{[2]}$ Shell | | | $G^{[3]}$ Shell | | | | |
| | | | Distance Quantum $a_*^{[-2]}$ = = 0.00329 $R_\odot$ | Distance Quantum $a_*^{[-1]}$ = = 0.044268 $R_\odot$ | Distance Quantum $a_*^{[1]}$=16.735 km | | | Distance Quantum $a_*^{[2]}$ = 225 km | | | Distance Quantum $a_*^{[2]}$ = 2.765 km | | | Distance Quantum $a_*^{[3]}$ = 37.174 km | | | | |
| | | | Radius $a=a_*^{[-2]} a^\wedge$ | Radius $a=a_*^{[-1]} a^\wedge$ | Radius $a=a_*^{[1]} a^\wedge$ | | Depth N=R-a | Radius $a=a_*^{[2]} a^\wedge$ | | Depth N=R-a | Radius $a=a_*^{[2]} a^\wedge$ | | Depth N=R-a | Radius $a=a_*^{[3]} a^\wedge$ | | Depth N=R-a | | |
| | | | [a/$R_\odot$] | [a/$R_\odot$] | [km] | [a/R] | [km] | [km] | [a/R] | [km] | [km] | [a/R] | [km] | [km] | [a/R] | [km] | | |
| ME | 8.083 | 10.398 | 0.0342 | 0.460 | 174 | 0.0272 | 6197 | 2430 | 0.366 | 4031 | 29 | 0.0165 | 1709 | 386 | 0.222 | 1352 | | |
| TR | 9.191 | 13.444 | 0.0442 | 0.595 | 225 | 0.0352 | 6146 | 3025 | 0.474 | 3346 | 37 | 0.0213 | 1701 | 500 | 0.287 | 1238 | | |
| V | 11.050 | 19.433 | 0.0639 | 0.860 | 325 | 0.051 | 6046 | 4372 | 0.685 | 1999 | 54 | 0.0309 | 1684 | 722 | 0.415 | 1016 | | |
| | | | | 1 | | | | | | | | | | | | | | |
| E | 12.993 | 26.868 | 0.0884 | 1.189 | 450 | 0.0705 | 5921 | 6045 | 0.947 | 326 | 74 | 0.0427 | 1664 | 999 | 0.574 | 739 | | |
| | | | | | | | | | 1 | 0 | | | | | | | | |
| (U) | 15.512 | 38.296 | 0.126 | | 641 | 0.100 | 5730 | | 1.344 | | 106 | 0.0609 | 1632 | 1424 | 0.819 | 314 | | |
| MA | 16.038 | 40.937 | 0.134 | | 685 | 0.107 | 5686 | | | | 113 | 0.065 | 1625 | 1522 | 0.875 | 216 | | |
| | | | | | | | | | | | | | | | 1 | 0 | | |
| (NE) | 19.431 | 60.091 | 0.197 | | 1005 | 0.157 | 5366 | | | | 166 | 0.095 | 1572 | 1738 | 1.277 | | | |
| CE | 21.614 | 74.351 | 0.244 | | 1244 | 0.195 | 5127 | | | | 205 | 0.118 | 1533 | | | | | |
| (P) | 22.235 | 78.685 | 0.259 | | 1317 | 0.206 | 5054 | | | | 217 | 0.125 | 1521 | | | | | |

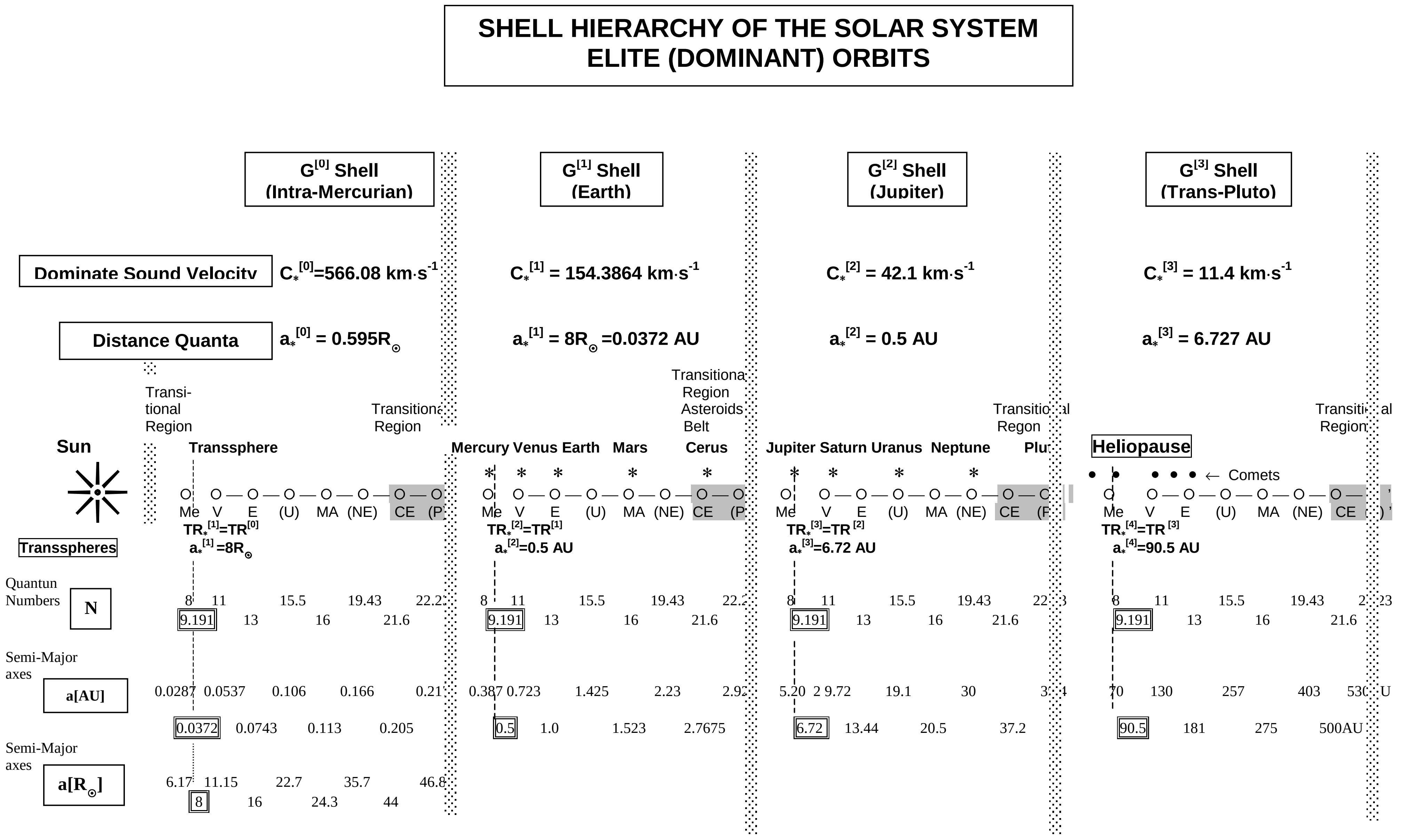
SHELL HIERARCHY OF THE SOLAR SYSTEM
ELITE (DOMINANT) ORBITS
G[0] Shell (Intra-Mercurian)
G[1] Shell (Earth)
G[2] Shell (Jupiter)
G[3] Shell (Trans-Pluto)
Dominate Sound Velocity
C*[0]=566.08 km·s-1
C*[1] = 154.3864 km·s-1
C*[2] = 42.1 km·s-1
C*[3] = 11.4 km·s-1
Distance Quanta
a*[0] = 0.595R⊙
a*[1] = 8R⊙ =0.0372 AU
a*[2] = 0.5 AU
a*[3] = 6.727 AU
Transi- tional Region
Transitional Region
Transitional Region Asteroids Belt
Transitional Regon
Transitional Region
Sun
Transsphere
Mercury Venus Earth Mars Cerus
Jupiter Saturn Uranus Neptune Plu
Heliopause
← Comets
Me V E (U) MA (NE) CE (P
TR*[1]=TR[0]
a*[1] =8R⊙
TR*[2]=TR[1]
a*[2]=0.5 AU
TR*[3]=TR [2]
a*[3]=6.72 AU
TR*[4]=TR [3]
a*[4]=90.5 AU
Transspheres
Quantun Numbers
N
8 11 15.5 19.43 22.2
9.191 13 16 21.6
8 11 15.5 19.43 22.2
9.191 13 16 21.6
8 11 15.5 19.43 22
9.191 13 16 21.6
8 11 15.5 19.43 23
9.191 13 16 21.6
Semi-Major axes
a[AU]
0.0287 0.0537 0.106 0.166 0.21
0.0372 0.0743 0.113 0.205
0.387 0.723 1.425 2.23 2.92
0.5 1.0 1.523 2.7675
5.20 9.72 19.1 30
6.72 13.44 20.5 37.2
70 130 257 403 530
90.5 181 275 500AU
Semi-Major axes
a[R⊙]
6.17 11.15 22.7 35.7 46.8
8 16 24.3 44

**Fig. 6**